\documentclass[12pt]{iopart}
\usepackage[dvips]{graphicx}

\begin{document}

\title[Band offset determination of the GaAs/GaAsN interface using the DFT method]
{Band offset determination of the GaAs/GaAsN interface using the DFT method}

\author{H-P Komsa$^1$, E Arola$^1$, E Larkins$^2$, T T Rantala$^1$}

\address{Department of Physics,
Tampere University of Technology,
Finland}
\address{Faculty of Engineering,
University of Nottingham,
United Kingdom}
\ead{hannu.komsa@tut.fi}
\begin{abstract}
The GaAs/GaAsN interface band offset is calculated from
first principles.
The electrostatic potential at the core regions of the atoms is used to
estimate the interface potential and align the band structures
obtained from respective bulk calculations.
First, it is shown that the present method performs well on the
well-known
conventional/conventional AlAs/GaAs (001)
superlattice system.
Then the method is applied to a more challenging
nonconventional/conventional GaAsN/GaAs (001)
system, and consequently type I band lineup and
valence-band offset of about 35 meV is obtained
for nitrogen concentration of about 3 \%, 
in agreement with the recent experiments.
We also investigate the effect of strain on the band lineup.
For the GaAsN layer longitudinally strained to
the GaAs lattice constant, the type II lineup with
a nearly vanishing band offset is found,
suggesting that the anisotropic strain
along the interface is the principal cause
for the often observed type I lineup.
\end{abstract}

\submitto{\JPCM}
\maketitle

\section{Introduction}

% 1. About band offsets (->motivation)
% 2. History review (GaAsN band offset review -> motivation)
Since the birth of the band-gap engineering
the estimation of the band offsets at interfaces has
become of utmost importance.
Control of band offsets enables the control of
charge carrier flow and confinement, which is
a basic requirement in practically all semiconductor
device design.
%eero:
%[Explain why GaAsN is interesting material? EL]
Unlike the conventional III-IV semiconductor alloys, whose
physical properties change smoothly as a function of the alloying
composition, GaAs$_{1-x}$N$_x$ alloys have attracted a plenty of
experimental and theoretical attention due to their unusual
physical properties.
For example, the incorporation of nitrogen into GaAs drastically reduces
the band gap, \cite{Weye92,Wei96} which makes this material technologically
attractive for optoelectronic devices. \cite{Semi02}
This feature, together with a peculiar electronic states
localization and/or delocalization behaviour near the band edges
of GaAsN, \cite{Kent01a} have obviously important implications to the
band offset properties of the GaAsN/GaAs heterostructure.

Recently, there has been debate on whether GaAsN grown on the
GaAs substrate would show a type I \cite{kris1,zhan2,klar1,buya2,egor1}
or type II \cite{kita1,sun1,chen1} band lineup,
i.e., whether the GaAsN valence-band maximum (VBM) is above or below
the GaAs VBM, respectively,
as both results have been obtained experimentally. Apart from the
early dielectric model \cite{saka1}, all the computational results
support type I lineup \cite{lind1,wu1,bell1,Gued07}.
Naturally, whether the band lineup is type I or type II has
a dramatic effect on the device performance.
Nowadays, it seems to be accepted that the lineup is type I.
After all, the devices designed, while expecting a type I lineup,
seem to be working.
However, the actual amount of the valence-band offset is still unknown, but
usually a few tens of meV is assumed. This means that the
band gap difference is almost completely on the
conduction-band offset.

% 3. Computational method review -> problems
Band offsets at the interfaces of heterostructures have been
calculated with more or less refined methods.
Previously, the methods usually concentrated on finding one
bulk-specific parameter, which could then be used
to determine the lineup. Most important methods
are the Anderson affinity rule \cite{ande1} and
Tersoff's theory of effective midgap states \cite{ters1}.
Attempts to determine the VBM in a global energy reference
can also be counted to this group.
These methods use no information about the interface, and
therefore lead to
the transitivity law for the valence-band offsets between
compound semiconductors as described by Wei and Zunger \cite{wei1}.
This transitivity property
holds fairly well for unstrained ``natural''
band offsets between pure semiconductor compounds \cite{wei1},
but fails in the case of strained layers.
A straightforward method would be to calculate complete
heterostructure and look at the local densities of states \cite{bass1},
but is practically troublesome.
Later on, methods that combine information
about the interface from one calculation and information about
the two bulk constituent systems
from separate calculations, have gained popularity
\cite{byla1,byla2,bald1,zhan1}.
%model solid approaches \cite{vand1} (pretty much same as interface dipole with TB),
This method combines the physically valid
framework with the computational affordability,
which has made it the most used method,
and will be our choice, too.
These methods can be performed using either
semi-empirical or {\it ab initio} computational framework.
Along with the abundance of computational resources it
has become possible to calculate band offsets
from first principles
at the interface of binary systems \cite{byla1,byla2,bald1,wei1}
and even some tertiary systems \cite{zhan1}.

Despite the general interest towards
the GaAs/GaAsN material system,
the computational considerations are hindered by the
low concentration of (randomly substituted) nitrogen
in the GaAsN layer, and the polymorphic nature of GaAsN,
resulting in the need of a large periodic computational cell.
For this reason, previous calculations have been using either
\mbox{{\boldmath $k$} $\cdot$ {\boldmath$p$}} method \cite{wu1},
tight-binding results fitted to the
BAC method \cite{lind1}, or
semiempirical pseudopotential methods \cite{bell1}. Purely
\emph{first-principles}
calculations have been missing until now.
Other notable differences are that in all of these approaches
the VBM energies of the bulk material constituents are
calculated and these do not consider interface effects,
and also the effect of strain due to lattice-matching
to the substrate has not been explicitly examined.
To solve these problems, we not only need to use large supercells, but also an
alternative way to extract the strength of the interface potential.

% 4. Article outline -> how we solve the problems
In this study, we present a revision of the old models modified
in a way that the electrostatic potential is determined from the
atomic cores.
The central part of our model is similar to the one
used by Jaffe \emph{et al.} \cite{Jaff04} in their
$\alpha$-Cr$_2$O$_3$/$\alpha$-Fe$_2$O$_3$ (0001) interface
band offset calculations, containing analogous features from
the experimental band-offset measurement technique based on
the core-state (x-ray) photoelectron spectroscopy, see, for example,
Ref.\ \cite{krau1}.
This should naturally alleviate the problems mentioned
above, and we will show in the following that this is the case.
In section 3 we demonstrate the applicability of our method on the
well-known AlAs/GaAs system. In addition to the band-offset values,
we provide additional insight into the electronic structure of the
states near the VBM by looking at the wavefunction localization.
Finally, we proceed on to the more challenging case of the
GaAsN/GaAs system and carry out a similar analysis there.

\section{Models and Methods}

% 1. Physics
In a semiconductor interface, lining up the Fermi-levels causes
charge accumulation at the interface, which creates
an electric field and subsequently a potential at the interface.
Since the energy reference of the band structure
can be related to the average
electrostatic potential, it is sufficient to estimate the change
in the average electrostatic potential through the interface
from a heterostructure calculation, and then align
the bulk valence-band maxima accordingly to obtain the
valence-band offset.

There are several ways to incorporate the concept of the average
electrostatic potential into the calculation of the band offset.
Bylander and Kleinman \cite{byla1,byla2} used a method where
the interfacial double-layer potential ($\Delta V$),
induced by the planar average of the difference between the superlattice
and the bulk constituent charge densities ($\Delta\rho$),
was used along with the bulk constituent $\Gamma_8$ eigenvalues
of the valence-band maximum, in order to determine the valence-band offset.
In the method of Baldereschi \emph{et al.} \cite{bald1}
running average across the unit cell,
along the growth direction,
of the $xy$ planar averaged potential
is calculated to obtain
a slowly varying curve for the potential, from where it is a simple thing
to read the potential shifts.
In principle, one could also take only one point (any point)
from the electrostatic
potential and compare that between the interface calculation
and the bulk calculation. In practice, due to charge redistribution
and geometric relaxations, a well-defined reference
point in space should be chosen.
An example of such a point would be
the potential in the core regions of the ions.

% Hannu
Merely from this reasoning, it is possible to write out
the equation for the band offset between the materials X and Y
\begin{eqnarray}\label{eq:potshift}
\Delta E_v
&
%= E_v[X]-E_v[Y]
= (E_v[X]-V_c[X])^b-(E_v[Y]-V_c[Y])^b + (V_c[X]-V_c[Y])^i \nonumber \\
&= (E_v[X]-E_v[Y])^b + (V_c[X]^i-V_c[X]^b-V_c[Y]^i+V_c[Y]^b) \nonumber \\
&= \Delta E_v^b + \Delta V_c^i
,
\end{eqnarray}

where $E_v$ is the valence-band maximum and
$V_{c}[X(Y)]$ is the electrostatic potential at the core
of a given type of an atom (anion or cation) located in the
material $X(Y)$.
Superscripts $b$ and $i$ stand for the bulk and interface
calculation, respectively.
Finally, $\Delta E_v^b$ is the lineup of the
VBM between the bulk constituents $X$ and $Y$, and
$\Delta V_c^i$ aligns the energy reference by combining
the electrostatic potentials at the core of some
appropriately chosen anions or cations
from the bulk and interface calculations.

Further justification for the use of the electrostatic potential
at the atomic core, in aligning the two bulk band structures
can be found from the idea of the band offset determination by
the core-level photoemission spectroscopy \cite{krau1}.
In this approach,
the core levels are measured with respect to the valence-band maxima.
At the same time, the change in the
core-level energy from one side of the interface (X)
to the other (Y) is also measured, so that the
valence-band offset $\Delta E_v$ at the heterojunction interface
is simply given by
\begin{equation}\label{eq:coreshift}
\Delta E_v = (E_v[X]-E_{cl}[X])-(E_v[Y]-E_{cl}[Y])+(E_{cl}[X]^i-E_{cl}[Y]^i),
\end{equation}

where $E_v[X(Y)]$ and $E_{cl}[X(Y)]$ are the valence-band maximum and
core-level energy of the material X(Y), far away from the interface,
and $E_{cl}[X(Y)]^i$ is the core-level energy at the X(Y) side
of the interface.
On the basis of our calculations for the band offset at the interface
of the AlAs/GaAs system (see Sec. 3.1 for details),
it seems obvious that in the common-anion system the anion core
potential can be used to accurately calculate the macroscopic average,
as defined by Baldereschi \emph{et al.} \cite{bald1},
of the electrostatic potential across the interface, and
therefore to accurately estimate the valence-band offset.
Similarly, a cation-related core potential can be used
in band-offset calculations for common-cation systems.

% 2. methods

Using the core-level energies in determining the band offset is
computationally problematic, since this approach requires,
if not an all-electron method, at least a method
with deep core potentials and subsequently sufficient number of valence
electrons active in the calculation.
A comprehensive list of band offsets for binary materials
calculated with this approach,
using the linearized augmented plane wave (LAPW) method,
has been given by Wei and Zunger. \cite{wei1}
For large systems, such methods are computationally
prohibitively expensive, and therefore pseudopotential or
frozen-core methods (e.g. the projector-augmented wave method, PAW)
would be more suitable ones.
The electrostatic potential at each ion (``the core potential'')
is calculated by placing a test charge at each ion and
calculating in the usual way
\begin{equation}
V_c({\bf R}_n) = \int{V({\bf r})\rho_{test}(|{\bf r}-{\bf R}_n|)d^3{\bf r}},
\label{corepot}
\end{equation}

i.e., we estimate
the test-charge distribution weighted average of
the electrostatic potential, a typical core electron 
would experience. More precisely, the integral is calculated
over a spatially extended spherically symmetric region
whose radius is related to the PAW core radius.
The norm of the test charge distribution is constrained to one.
The radii for the test charge distributions are taken
1.03 {\AA} for gallium, 1.06 {\AA} for arsenic and 0.75 {\AA} for nitrogen.

Although this method can be applied
to any material system, with or without a lattice-matched interface,
it is especially useful in calculating the band offset of
the nitrogen dilute and low-concentration GaAsN alloy
with respect to GaAs, as can be seen from the following points.
First of all, a laterally wide computational cell is required
for these nitride material systems, but, on the other hand,
it is known that the
macroscopic average of the electric dipole-related
electrostatic potential becomes saturated rapidly away from
the interface, and therefore only a few layers of the material is
needed in the longitudinal direction to estimate the interface potential.
Second, the calculation of the running average of the planar-averaged
electrostatic potential over conventional 8-atom unit cells
is ill-defined since in the dilute and low-concentration
nitride systems the atomic positions around the nitrogen atoms are
strongly displaced from their ideal
positions in the zinc-blende lattice structure.
Third, the valence-band offset in the GaAsN/GaAs systems
is known to be very small, so reading the band offset from the
layer-projected local density of states (LDOS) is practically impossible,
particularly due to the relatively small thickness of these layers.
Finally, the method used for the band offset calculation for
the AlAs/GaAs system in Refs.\ \cite{byla1} and \cite{byla2},
requires that the planar averages of both the bulk and
superlattice electrostatic potentials have been computed
using the same {\bf k}-point mesh, and that identical
lattice geometries have been used in the both cases.

% 3. computational details

In our studies presented here,
all calculations were performed within the density functional theory
(DFT) framework using a plane-wave basis as implemented in
the code {\scshape VASP}
(Vienna {\it ab initio} simulation package) \cite{kres1,kres2,kres3}.
The atomic core regions are described using
projector-augmented waves (PAW) with the $3d$ electrons of gallium
explicitly included in the calculation. The local density approximation (LDA)
of Ceperley and Alder is used, as parametrized by Perdew and Zunger
\cite{cepe1,perd1}.

The spin-orbit coupling (SOC), making the computation a lot more
time consuming, is not included in our band-offset calculations.
However, the following points obviously justify us omitting the SOC from
our calculations.
Firstly, on the basis of our band-offset test calculations for the
AlAs/GaAs system, using a supercell composed of $4 + 4$ unit cells,
it seems that there is hardly any change in the Coulomb potential
behaviour across the interface when the SOC is included
(e.g. the planar averaged quantity $\Delta V_C^i$ of Eq.\ (\ref{eq:potshift})
changes only about 1 meV when the SOC is switched on).
Secondly, our test calculations on the AlAs/GaAs system show that the
valence-band offset increases by some 13 meV when the SOC is included.
It is noticeable that this change in the band offset is only about one
third of the difference between the spin-orbit (SO) -splitting at the
VBM ($\Delta_0$) in GaAs (0.34 eV) and in AlAs (0.30 eV).
Finally, the electroreflectance experiments by Perkins \emph{et al.}
\cite{per99} for GaAs$_{1-x}$N$_x$ alloys reveal that the SO-splitting
at the VBM stays nearly constant for $0 < x < 3\%$, being 0.33 eV
for $x=2.2\%$. Also, our calculations with the SOC
give very similar $\Delta_0$ values of 0.33 eV and 0.31 eV for the isotropic
bulk GaAs$_{1-x}$N$_x$ ($x=3.125\%$) and for GaAs$_{1-x}$N$_x$ ($x=3.125\%$)
laterally matched with GaAs, respectively.
Given that the abovementioned \emph{ad hoc}
``one third''-rule can be transferred
also to the GaAsN/GaAs system, we can conclude that the omission of the
SOC from our calculations causes less than  (0.34 eV - 0.31 eV)/3 = 10 meV
error to the band-offset result for the GaAsN/GaAs interface \cite{soc_strain}.

Since the DFT works reliably for the ground state and the LDA is known to
seriously underestimate the fundamental band gap, values of
the conduction-band offset from these calculations
can not be expected to be reliable.
However, the knowledge of the valence-band offset makes it possible to estimate
the conduction-band offset if the band gaps are experimentally known.

\section{Results}

We begin by testing our method on the well-known AlAs/GaAs
material system. We compare the results from
two different methods.
In the first method, the interface potential has been estimated
using the electrostatic potential at the atomic cores
[see Eq.\ (\ref{corepot})], and in the second one
using the average electrostatic potential technique of
Ref.\ \cite{bald1}.
Naturally, comparison to previous results in the literature
is also given.

\subsection{AlAs/GaAs interface}

A good review of the experimental data of the band offset for the
AlAs/GaAs system is given by Vurgaftman \emph{et al.} \cite{vurg1}.
A split of 65:35 between conduction- and valence-band discontinuities
is generally agreed upon. The direct band gap for GaAs is 1.424 eV
and for AlAs about 2.95 eV,
%(from valence-band maximum to $\Gamma$-point in conduction band),
resulting in a valence-band offset of about 0.53 eV at 300 K.
These results show no significant differences between
the 0 K and 300 K temperatures.
The LAPW result obtained by Wei is 0.51 eV \cite{wei1}.

Our computational cell is 16 cubic unit cells long and one unit cell
wide in the two lateral directions, resulting in a total of 128 atoms.
The first 8 unit cells form the AlAs layer and the following 8
unit cells form the GaAs layer.
The lattice constant is chosen so that the system is lattice matched
to GaAs in all directions. Even though the supercell
size has not been optimized, the atomic geometry within the cell
has been relaxed.
As the computationally relaxed GaAs lattice constant is 5.605 \AA, this
corresponds to layers of about 4.5 nm
thickness in the $\langle 100\rangle$ direction.
This is in the limit of
experimentally achievable quantum well widths.
4x4x2 Monkhorst-Pack {\bf k}-point set is used to calculate the total energy
and atomic geometry relaxation, and a 9x9x3 {\bf k}-point
set is used to calculate the density of states.

\begin{figure}[t]
\begin{center}
  \includegraphics[width=10cm]{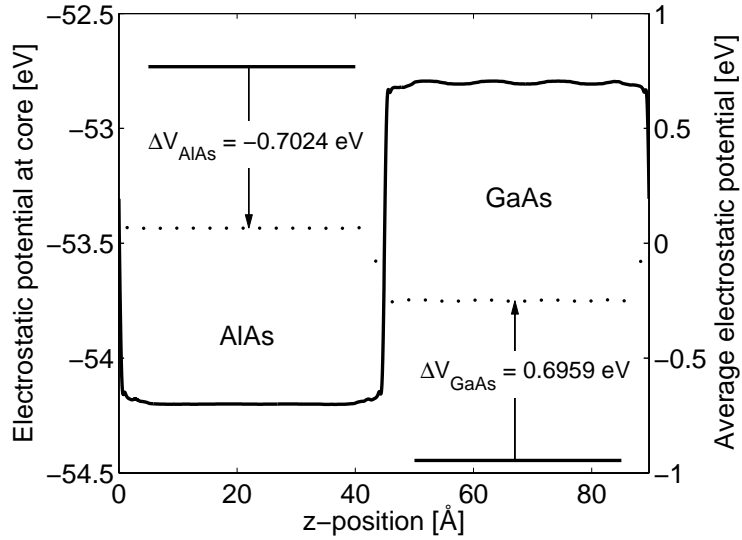}
\end{center}
\caption{\label{fig:AlGaAspots}
Electrostatic potential at the cores of arsenic atoms
in the supercell of the AlAs/GaAs system (dotted line),
and the corresponding values in the AlAs and GaAs bulk constituents
(solid horizontal lines) [energy scale on the left].
Also shown is the macroscopically averaged electrostatic potential
(solid line, see Ref.\ \cite{bald1}) where the averaging
steps along the longitudinal direction of the supercell have been
carried out over the 8-atom unit cells [energy scale on the right].
}
\end{figure}

First, we consider our method of lining up the electrostatic potentials
at the cores of arsenic atoms.
In \fref{fig:AlGaAspots} the potential at the arsenic core
is plotted along the longitudinal direction of
the supercell together with the corresponding bulk values.
As compared to the bulk, in the heterostructure calculation,
the arsenic core potential on the GaAs side is increased
by $0.6959$ eV and on the AlAs side it is decreased by
$-0.7024$ eV, resulting in 
$\Delta V_c^i = 1.3983$ eV.
The valence-band maximum is at 
$E_v[GaAs] = 3.1608$ eV in the GaAs bulk and 
$E_v[AlAs] = 4.0412$ eV in the AlAs bulk,
Eq.\ \ref{eq:potshift} giving out
the band offset $\Delta E_v = 0.5179$ eV.

Next, we calculate the macroscopically averaged electrostatic
potential along the longitudinal direction of the
supercell following the method of
Baldereschi \cite{bald1}.
The result for this is also shown in \fref{fig:AlGaAspots}.
The average over the averaged macroscopical
potential is $\langle\bar{V}[GaAs]\rangle = 0.7000$ eV on the GaAs side
of the supercell and
$\langle\bar{V}[AlAs]\rangle = -0.7003$ eV on the AlAs side,
yielding almost the same interface potential of
%eero: In Baldereschi's method there should be no c ("core") subscript.
$\Delta V^i = 1.4002$ eV.
The average of the electrostatic potential of the bulk is zero
[in this case $\Delta V^i$ corresponds to
$\Delta V_c^i$ of Eq.\ (\ref{eq:potshift})],
so we can readily shift the bulk VBM values with this, yielding
$\Delta E_v = 0.5198$ eV. These results are very close to
the experimental values and that from the LAPW method.
It is clear that the two methods produce very similar results
and that our ``core level'' method appears to be well justified.

At this point, we briefly
mention that the difference
between our computational lattice constants of AlAs ($a=5.635$ \AA)
and GaAs ($a=5.605$ \AA) is 0.03 {\AA}, while the corresponding
experimental difference is only 0.008 {\AA}.
This results in a slightly overestimated strain of the AlAs/GaAs
system in our calculations.

It has been shown in the previous studies that only a few
atomic layers of material is sufficient for the potential to
become saturated away from the interface (cf. e.g. Ref.\ \cite{bald1}).
This is also evident from \fref{fig:AlGaAspots}.
This turns out to be a case, also in the GaAsN/GaAs system
(see \sref{ssec:GaAsNif}).

\begin{figure}[t]
\begin{center}
  \includegraphics[width=10cm]{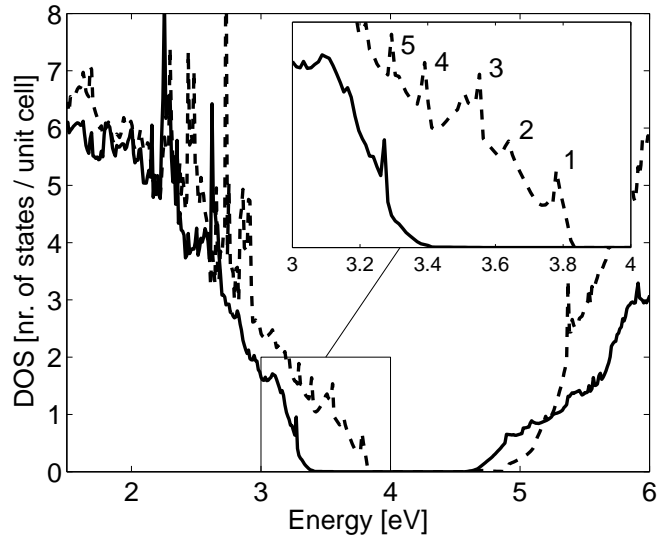}
\end{center}
\caption{\label{fig:AlGaAsLDOS}
The local DOS of one 8-atom unit-cell wide slice from the
AlAs layer (solid line) and GaAs layer (dashed line)
of the AlAs/GaAs (001) $1\times 1\times 16$ supercell.
}
\end{figure}

\begin{figure}[t]
\begin{center}
  \includegraphics[width=10cm]{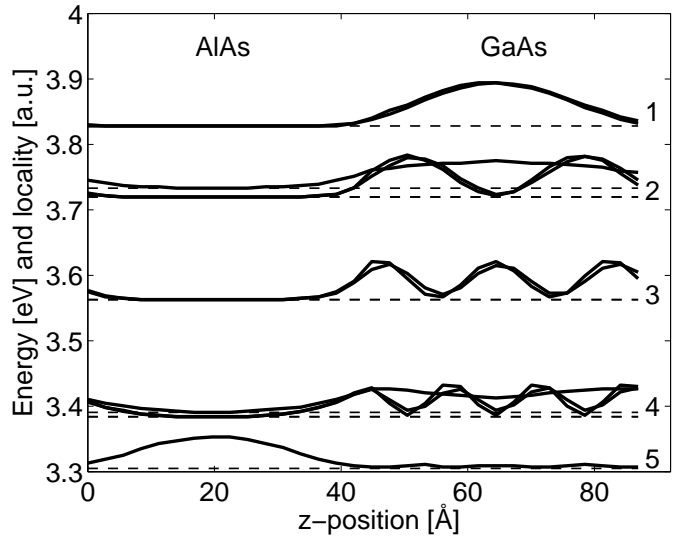}
\end{center}
\caption{\label{fig:AlGaAsstates}
Envelope-like functions
of the highest valence-band states at the $\Gamma$ point,
showing the localization of these states.
Dashed lines show the energy levels.
Envelope-like functions are in arbitrary units,
whose zero level has been defined by their corresponding
energy levels.
}
\end{figure}

In order to justify the analysis of the wavefunctions
on the GaAs/GaAsN system in the next section, we take
a closer look at the one-electron states
in the AlAs/GaAs (001) superlattice system.
Firstly, in \fref{fig:AlGaAsLDOS}, we show the local DOS
projected onto one 8-atom unit-cell wide slice
taken in the middle of the AlAs and GaAs layers.
In principle, one could also estimate the band offset from
this figure (around 0.5 eV). However, it is
difficult to extract precise values, and to be exact, this is
not the band offset, but the difference between the highest
quantum well (QW) confined state and the highest unconfined state.
This value is smaller than the band offset but it
should approach the band offset as the width of the QW
increases. However, this is an impractical method, because of the
prohibitively large computational volume required.
Furthermore, in \fref{fig:AlGaAsstates}, we show the
envelope-like functions \cite{env_funct}
of the highest confined hole states at the $\Gamma$ point.
To be more precise, the localization of the
superlattice wavefunction (Bloch wave), projected
onto the slice of half the 8-atom unit cell,
is plotted along the longitudinal direction of the supercell
and vertically shifted to the corresponding energy eigenvalue.
It is clear, that this slice-projected localization of
the wavefunction in one region of a heterostructure
means confinement of the charge carriers at that region,
and here particularly confinement
of holes due to the QW structure.
The five LDOS peaks in the inset of \fref{fig:AlGaAsLDOS},
are derived from the centre-slice of the GaAs slab, and
can obviously be related to the confined mini-band states of the
AlAs/GaAs (001) superlattice structure. 

\subsection{GaAsN/GaAs interface}\label{ssec:GaAsNif}

\begin{figure}[t]
\begin{center}
  \includegraphics[width=10cm]{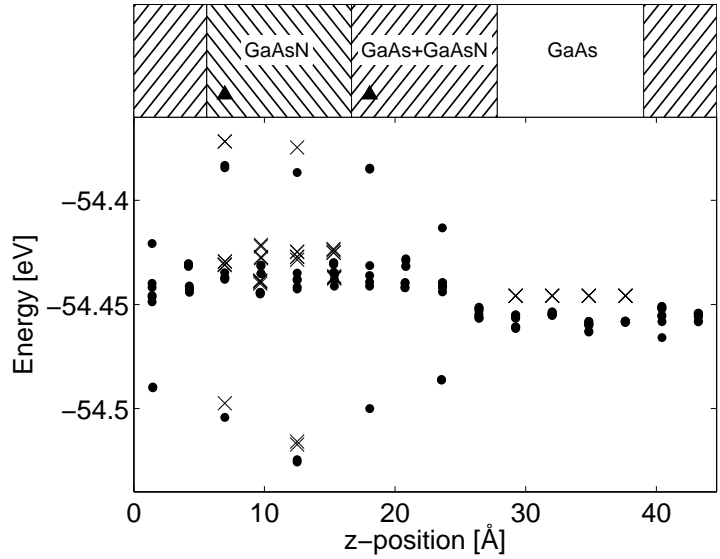}
\end{center}
\caption{\label{fig:GaAsNcore} Electrostatic potential
at cores of arsenic atoms along the longitudinal direction
of the GaAsN/GaAs (001) supercell.
The dots are from the superlattice calculation
and the crosses are from
the GaAsN and GaAs bulk constituent calculations.
On the top of the picture is depicted
a construction of the $2\times 2\times 8$ supercell.
Triangles show the positions of the
two nitrogen atoms in the supercell.
}
\end{figure}

The practical nitrogen concentration in GaAsN
is usually only a few percent due to the miscibility gap
restricting the growth.
If one arsenic is replaced by a nitrogen in a 64-atom
cubic supercell it corresponds to a nitrogen concentration of about 3 \%.
Therefore, our bulk calculations have been performed with a
64-atom supercell and using 2x2x2 {\bf k}-space sampling.
Since there is only one nitrogen in the 64-atom
supercell, for the minimal construction of the heterostructure
supercell we need two 64-atom
GaAsN supercells and preferably same amount of GaAs,
giving out a 256-atom supercell. Because the
GaAsN layers have to be lattice matched to GaAs in the
lateral directions, this also requires that the longitudinal
lattice constant has to be optimized to give the minimum total
energy. Since the 256-atom
supercell calculations are computationally
expensive we carry out the longitudinal lattice constant optimization
of the bulk and interface regions separately.
Interface effects on the cell size are estimated from a
128-atom supercell and 2x2x1 {\bf k} points.
When building a 256-atom heterostructure supercell,
we interlace the GaAsN and GaAs bulk geometries from
their 64-atom bulk calculations and GaAsN/GaAs (001)
interface geometries from the 128-atom calculation.
Construction of this supercell can be seen on the top of
\fref{fig:GaAsNcore}.
After this we perform one more relaxation for the ions inside the 256-atom
supercell without optimization of the longitudinal lattice constant.

Although it is difficult to say
what the ''effective nitrogen concentration'' of the
GaAsN layer in this supercell is, we assume
that the roughly 3 \% nitrogen concentration of
our GaAsN bulk calculation would be a good estimate.
It is worth pointing out that the real GaAsN alloy
has a substitutionally random nitrogen distribution,
while the GaAsN phase in our GaAsN/GaAs (001) superlattice
calculation repeats itself periodically,
and therefore defines an ordered distribution of N atoms.

The electrostatic potential at the cores of arsenic atoms
is shown in \fref{fig:GaAsNcore} for both interface
and bulk calculations. One can see
saturation, when moving away from the interface,
in the potential values, in both the GaAsN and GaAs layers.
The single isolated potential values below the mean are due to
the arsenic atoms between nitrogen atoms along the (110) directions.
Those above the mean are due to the arsenic atoms between nitrogen
atoms along the (100) directions.

While averaging over the
arsenic core potentials in the middle of
the GaAs and GaAsN regions, a suitable volume has to be selected.
We have tested three different volumes
on the GaAsN side of the supercell:
''mid1'' has only one monolayer of arsenic atoms between the nitrogen atoms,
''mid3'' has 3 monolayers of arsenic atoms between the nitrogen atoms,
and ''64c'' has 4 monolayers of arsenic atoms consisting of
the same arsenic atoms as in the original
64-atom GaAsN supercell that was embedded
into the heterostructure supercell.
In the same way, the averaging volumes ''mid1'', ''mid3'', and  ''64c'',
will be defined on the GaAs side of the supercell. 

\begin{table}
\caption{\label{tab:offsets}
Band offsets calculated with different averaging models.}
\begin{indented}
\item[]\begin{tabular}{@{}lllllll}
\br
\centre{7}{Longitudinally relaxed superlattice}\\
\ms
\crule{7}\\
\ms
      & \centre{2}{VBM} & \centre{2}{$V_c^i[X(Y)]-V_c^b[X(Y)]$} & & \\
\ns
      & \crule{2}       & \crule{2}              & & \\
model & GaAsN & GaAs    & $X\equiv$GaAsN & $Y\equiv$GaAs & $\Delta V_c^i$ & $\Delta E_v$ \\
\mr
mid1  & 3.1920 & 3.1608 & -0.0115 & -0.0143 & 0.0028 & 0.0340 \\
mid3  & 3.1920 & 3.1608 & -0.0074 & -0.0117 & 0.0043 & 0.0355 \\
64c   & 3.1920 & 3.1608 & -0.0076 & -0.0118 & 0.0042 & 0.0354 \\
\mr
\centre{7}{Longitudinally lattice-matched superlattice}\\
\ms
\crule{7}\\
\ms
      & \centre{2}{$VBM$} & \centre{2}{$V_c^i[X(Y)]-V_c^b[X(Y)]$} & & \\
\ns
      & \crule{2}       & \crule{2}              & & \\
model & GaAsN & GaAs    & $X\equiv$GaAsN & $Y\equiv$GaAs & $\Delta V_c^i$ & $\Delta E_v$ \\
\mr
mid1  & 3.0499 & 3.1608 & 0.0507 & -0.0576 & 0.1083 & -0.0026 \\
mid3  & 3.0499 & 3.1608 & 0.0500 & -0.0572 & 0.1072 & -0.0037 \\
64c   & 3.0499 & 3.1608 & 0.0489 & -0.0567 & 0.1056 & -0.0053 \\
\br
\end{tabular}
\end{indented}
\end{table}

Valence-band maxima, changes in the averaged
electrostatic potential, and finally the
band offsets are collected in \tref{tab:offsets}. We can see, that
using different averaging models,
has only a minor effect on the band offset.
A type I band offset of about 35 meV is obtained in the
longitudinally relaxed case.
Looking at the results
of \tref{tab:offsets}, for the longitudinally relaxed superlattice case,
one could even go as far as to claim that the 3 \% nitrogen concentration
is dilute enough such that $\Delta V_c^i$
could be approximated to be zero.
This assumption can obviously be transferred to calculate the band offset
only from the bulk VBM values
in the case of the GaAsN/GaAs superlattice containing an even more
nitrogen-dilute GaAsN layer.
We have carried out a one-{\bf k}-point ($\Gamma$)
calculation for a 216-atom cubic supercell
(composed of $3\times 3\times 3$ conventional
8-atom GaAs unit cells and one nitrogen atom) lattice-matched
to the GaAs bulk in two lateral directions (along the interface)
and relaxed in the third one (perpendicular to the interface).
In this case, the isotropic nitrogen concentration, being 0.93 \%
in the GaAsN layer, leads to the VBM value of 3.1673 eV and the
valence-band offset of about 7 meV.

Although the GaAs bulk VBM is located at $E_v[GaAs] = 3.1608$ eV at
the $\Gamma$ point, and is triply degenerate (when the SOC not included),
the corresponding eigenvalues for the GaAsN bulk with SOC
have about 48 meV split due to the anisotropic strain.
Therefore, even if the upper GaAsN valence band
is located higher than the GaAs VBM, and the type I
band offset is observed, the lower GaAsN valence band
is located at slightly lower energy than the GaAs VBM,
suggesting only a weak charge carrier confinement in
the GaAsN layer of the GaAsN/GaAs (001) superlattice.

\begin{figure}[t]
\begin{center}
  \includegraphics[width=10cm]{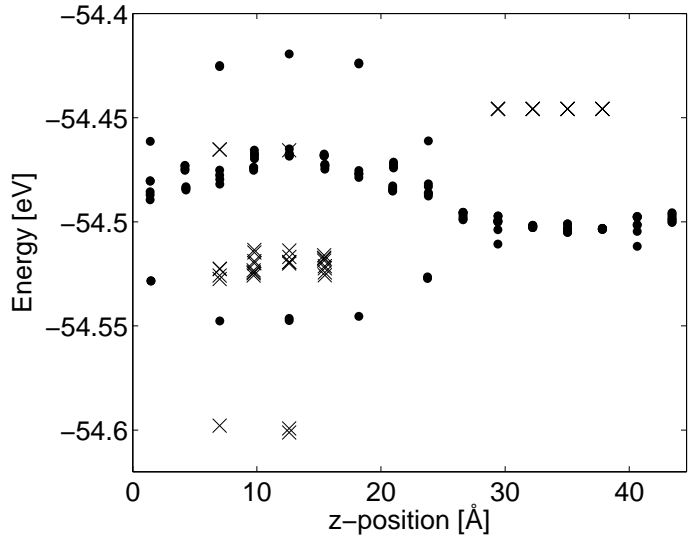}
\end{center}
\caption{\label{fig:GaAsNcorestrain} Electrostatic potential
at the cores of arsenic atoms along the longitudinal direction
of the GaAsN/GaAs (001)
longitudinally lattice-matched supercell.
The dots are from the interface calculation
and the crosses are from the
GaAsN and GaAs bulk constituent calculations.
}
\end{figure}

We also tested what would happen if the anisotropic strain is removed.
For this purpose we longitudinally stretched the heterostructure supercell to
match the GaAs lattice constant, although unfortunately this still leaves
the isotropic strain component left.
The data of the longitudinally lattice-matched case is also shown in
\tref{tab:offsets} and a picture of the electrostatic potential distribution
at the cores of arsenic atoms is shown in \fref{fig:GaAsNcorestrain}.
In this case the band offset is changed
to type II, but with a nearly vanishing band offset of about 4 meV.

Our results are in agreement with the experimental
results of Egorov \cite{egor1} and also with the
computational results of Lindsay \cite{lind1} and Bellaiche \cite{bell1}.
Egorov gives the band offset of $15\pm5$ meV
for the nitrogen concentration of about 2 \%.
By applying linear interpolation onto this experimental value
we get about 25 meV for the 3 \% concentration
and about 7 meV for the 1 \% concentration.
From the data of Bellaiche the band offset of
about 20 -- 30 meV for the case of 3 \% concentration,
can be estimated in good agreement with our results.
Finally, Lindsay gives a formula for the VBM evolution as $E_v(x) = 0.2x$ eV,
and therefore for the case of the 3 \% concentration,
this means about 6 meV value for the band offset, providing that
the contribution of the interfacial Coulomb potential change
$\Delta V_c^i$ could be ignored.
Additionally, Egorov also presents a band-offset estimate
for an unstrained case,
which is $0\pm5$ meV (for the 2 \% nitrogen concentration).
The tendency seen in our computational observations matches with this
experimental behaviour, and also with the recent tight-binding (TB)
empirical band offset calculations by Shtinkov \emph{et al.} \cite{Shti04}
for the GaAs$_{1-x}$N$_x$/GaAs (001) QW system,
in the case where the anisotropic strain was removed.

If the isotropically strained GaAsN layer
in the GaAsN/GaAs superlattice leads to a band offset
very close to zero, then our
results give a strong indication that the type I lineup of
this material system is mainly caused by
the VBM split states due to
the anisotropic strain of the GaAsN layer, and without
this strain there occurs no band offset,
or even a type II lineup can be expected in this situation.

\begin{figure}[t]
\begin{center}
  \includegraphics[width=10cm]{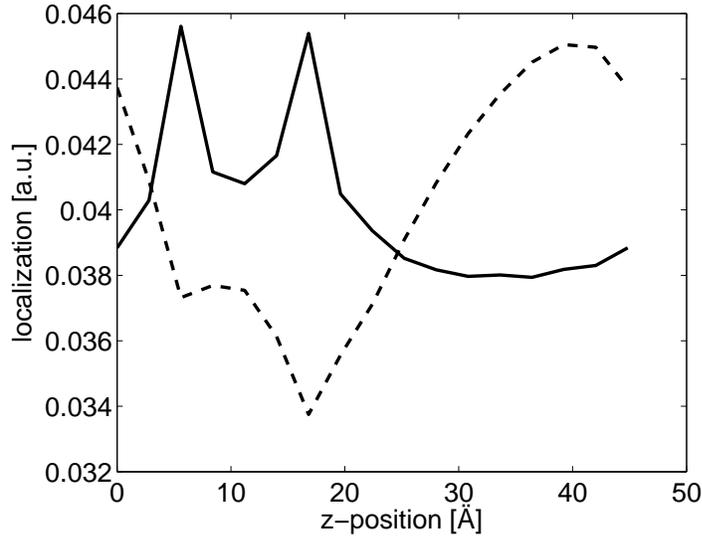}
\end{center}
\caption{\label{fig:GaAsNVBMloc} Localization of the
upmost valence-band states along
the longitudinal direction of the
the supercell, presented in
the same way as in \fref{fig:AlGaAsstates}.
The solid line is from the longitudinally relaxed superlattice
and the dashed line is from the longitudinally
lattice-matched superlattice.
}
\end{figure}

More insight can be gained by looking at the
wavefunction localization along the
longitudinal direction of the supercell for
the uppermost valence-band state (located at the $\Gamma$ point)
in both the longitudinally relaxed and the
longitudinally lattice-matched cases.
This is shown in \fref{fig:GaAsNVBMloc}.
In the case of the longitudinally relaxed supercell, 
this state is localized strongly on the nitrogen atoms, and
on the average residing on the GaAsN side, as is natural if
the GaAsN layer in the GaAsN/GaAs (001) superlattice
is supposed to feature type I lineup.
In the case of the longitudinally lattice-matched supercell,
the situation is the opposite.
The wavefunction is now localized on the GaAs side
as is natural for the type II lineup.
Interestingly, in this case charge is depleted from the nitrogen-localized
states region in the GaAsN layer of the supercell, into the GaAs layer.
This is directly reflected in the change of the interfacial Coulombic
potential term $\Delta V_c^i$ of Eq.\ (\ref{eq:potshift}),
which furthermore derives from the change in the induced electric dipole
across the interface.
Remarkably, it can be seen from \tref{tab:offsets} that this change in
$\Delta V_c^i$ is rather large, being more than 100 meV,
when going from the longitudinally relaxed superlattice to
longitudinally lattice-matched one.
Obviously, the electron charge localization behaviour at the nitrogen atoms,
much responsible for the change in $\Delta V_c^i$,
is a result of a subtle interplay between the QW-like confinement,
strain, and polymorphic nature of the GaAsN alloy.

In view of our computational approach,
we briefly comment in the following our band-offset calculations between
the AlAs/GaAs (001) and GaAsN/GaAs (001) superlattice cases.

Figure\ \ref{fig:AlGaAspots}, referring to the case of the
AlAs/GaAs (001) superlattice, clearly indicates that the
macroscopically averaged potential, as well as the core potential at
the As sites, change rapidly along the $z$ direction
near the interface,
%in the middle of the supercell, 
and then quickly saturate to an essentially
constant value. It is exactly this kind of behaviour in the electrostatic
potentials that can be used to define an "abrupt interface" between two
semiconductors, and typically exists between two ``conventional''
isoelectronic semiconductors (e.g. the GaAs$_{1-x}P_x$ alloy).

However, as Figures \ \ref{fig:GaAsNcore} and \ref{fig:GaAsNcorestrain}
show, the situation in the case of the GaAsN/GaAs (001) superlattice,
is quite the opposite. Due to the slow spread-out of the core potential
at As sites (in the $z$ direction) across the middle of the
supercell, the ``thickness'' and location of the interface,
from the electronic structure viewpoint,
are not so well-defined anymore.
Furthermore, obviously owing to the polymorphic nature of GaAsN alloy
and sensitive localization properties of the VBM states at the
nitrogen atoms in the GaAsN/GaAs superlattice,
there are rather large fluctuations at the core potential of the As sites.
However, the planar averaged quantity of the As core potentials
obviously leads to a smooth behaving curve along the $z$ direction,
with two well-defined plateaus
(one in the GaAsN layer, and the other one in the GaAs layer) which
can be used to determine $\Delta V_c^i$.

From these computational observations we can conclude that the concept
of the band offset may become ill-defined for the GaAsN/GaAs superlattice,
particularly in the case where the GaAsN layer is very thin.
%

%eero:
Finally, we briefly discuss the relationship between the atomic-
and electronic structural properties and optical properties
in GaAsN/GaAs systems.
First of all, the optical studies by Pan \emph{et al.} \cite{Pan99}
on GaAs$_{1-x}$N$_x$ layers on the GaAs (001) substrate and by
Gao \emph{et al.} \cite{Gao04} on the GaAsN/GaAs QW structure in connection
with the strain-compensating InAs layers
clearly demonstrated that optical properties,
in terms of photoluminescence (PL) emission,
are highly sensitive to the strain state and its relaxation of the GaAsN layer.
Secondly, Luo \emph{et al.} \cite{Luo03} observed a strong excitation energy-
and rapid thermal annealing (RTA) sensitive feature at the photon energy
of 1.385 eV (denoted as M in their paper) in the PL spectra of the
GaAsN/GaAs quantum well samples, which they interpreted to be due to
the interface-related localized exciton emission.
Interestingly, this PL feature can be greatly reduced by the RTA
treatment, therefore improving the ''optical quality`` of the
GaAsN/GaAs interface via reducing the localization traps at the interfaces.
Therefore, on the basis of our band offset calculations and these
experimental observations, we can conclude that the strain state of
the GaAsN layer in the GaAsN/GaAs QW largely influences both its band
offset (type of the band offset and its value) as well as PL properties.
Furthermore, it is obvious that the localized excitonic trap states
at the GaAsN/GaAs interface have effect on, not only on the PL spectra,
but also on the band offset itself.

\section{Conclusions}

We have presented an alternative way of extracting
the interface potential at a heterostructure interface from
the electrostatic potentials estimated
at the atomic cores. This method
is suitable for the systems with large geometric displacements
and not so well-defined interface, in terms of its location and
distribution along the growth direction, i.e. the interface abruptness.
After performing tests with the well-known AlAs/GaAs system
we have applied it to study the technologically interesting
GaAsN/GaAs interface.

We find a type I lineup with a band offset of about 35 meV for 
GaAsN layer of about 3 \% nitrogen concentration laterally lattice
matched to GaAs and about 7 meV for the 1 \% layer.
Moreover, a type II lineup with the
band offset close to zero is found for
a GaAsN layer strained to fit
the GaAs lattice constant in all directions.
Therefore, it is to be expected that the type I band offset
is largely due to the VBM split caused by the anisotropic strain.
We also studied the nitrogen localization by looking at the
envelope-like wavefunctions of the VBM states.

%eero:
We have recently started atomic and electronic structural studies on the
quinary GaInAsNX/GaAs material system
(X is some appropriate impurity additive),
which for X=Sb, has recently been demonstrated to be among the best
candidates for semiconductor lasers operating at wavelengths
longer than 1.3 $\mu$m \cite{Ishi06b,Harr05}.

\section*{Acknowledgments}

We are thankful for the Centre for Scientific Computing (CSC)
in Finland and
Material Sciences National Grid Infrastructure (M-grid, akaatti)
for the computational resources.

\section*{References}

\bibliography{gaasnbo}

\begin{thebibliography}{10}
\bibitem{soc_strain}
When estimating the correction of including the spin-orbit coupling (SOC), into
  the band offset at the interface, we have to notice that not only the
  SOC-split valence levels of the two separate semiconductor materials
  determine the band offset at their heterojunction, but also the strain at the
  interface will couple in a complicated manner to these individual SOC-split
  valence states.
\bibitem{env_funct}
The concept of the ``envelope-like function'' we use here, is not directly
  related to its standard definition of ``envelope function'' which appears in
  the envelope function approximation in the framework of the effective mass
  approximation for electrons and holes in the QW (for further details, see,
  for example, Chapter 9 of reference \cite{yu01}).
\end{thebibliography}

\end{document}